\documentclass[sigplan,10pt]{acmart}
\renewcommand\footnotetextcopyrightpermission[1]{}

\startPage{1}

\setcopyright{none}

\usepackage{graphicx}
\usepackage{booktabs}
\usepackage{xspace}
\usepackage{amsmath}
\usepackage{url}
\usepackage{listings}
\usepackage{xcolor}
\usepackage{hyperref}
\usepackage{subcaption}
\usepackage[capitalize]{cleveref}
\usepackage{enumitem}
\usepackage{siunitx}

\usepackage{algorithmicx}
\usepackage{algpseudocode}
\usepackage{algorithm}

\usepackage{booktabs}
\usepackage{tabularx}
\usepackage{makecell}
\usepackage{ragged2e}

\usepackage{multirow}

\newcommand{\KVC}{KV Cache\xspace}
\newcommand{\KVCs}{KV Caches\xspace}
\newcommand{\KVCc}{KV-Cache-centric\xspace}

\newcommand{\project}{TokenDance\xspace}

\newcommand{\AG}{All-Gather\xspace}

\newcommand{\KVCollector}{KV Collector\xspace}
\newcommand{\DStorer}{Diff-Aware Storage\xspace}

\newcommand{\GA}{GenerativeAgents\xspace}
\newcommand{\AS}{AgentSociety\xspace}

\begin{document}

\title{\project: Scaling Multi-Agent LLM Serving via Collective KV Cache Sharing}

\author{Zhuohang Bian}
\affiliation{
  \institution{Peking University}
  \city{Beijing}
  \country{China}
}
\email{22373017@buaa.edu.cn}

\author{Feiyang Wu}
\affiliation{
  \institution{Peking University}
  \city{Beijing}
  \country{China}
}
\email{2501111907@stu.pku.edu.cn}

\author{Chengrui Zhang}
\affiliation{
  \institution{Peking University}
  \city{Beijing}
  \country{China}
}
\email{2301111796@stu.pku.edu.cn}

\author{Hangcheng Dong}
\affiliation{
  \institution{Shanghai Jiao Tong University}
  \city{Shanghai}
  \country{China}
}
\email{donghangcheng@sjtu.edu.cn}

\author{Yun Liang}
\affiliation{
  \institution{Peking University}
  \city{Beijing}
  \country{China}
}
\email{ericlyun@pku.edu.cn}

\author{Youwei Zhuo}
\affiliation{
  \institution{Peking University}
  \city{Beijing}
  \country{China}
}
\email{youwei@pku.edu.cn}

\renewcommand\footnotetextcopyrightpermission[1]{}

\begin{abstract}
Multi-agent LLM applications organize execution in synchronized rounds where a central scheduler gathers outputs from all agents and redistributes the combined context.
This \emph{\AG} communication pattern creates massive \KVC redundancy, because every agent's prompt contains the same shared output blocks, yet existing reuse methods fail to exploit it efficiently.

We present \project, a system that scales the number of concurrent agents by exploiting the \AG pattern for collective \KVC sharing.
\project's \KVCollector performs \KVC reuse over the full round in one collective step, so the cost of reusing a shared block is paid once regardless of agent count.
Its \DStorer encodes sibling caches as block-sparse diffs against a single master copy, achieving $11$--$17\times$ compression on representative workloads.
Evaluation on \GA and \AS shows that \project supports up to $2.7\times$ more concurrent agents than vLLM with prefix caching under SLO requirement, reduces per-agent \KVC storage by up to $17.5\times$, and achieves up to $1.9\times$ prefill speedup over per-request position-independent caching.

\end{abstract}

\maketitle

\section{Introduction}
\label{sec:intro}

Large language models (LLMs) are the execution engine of many multi-agent platforms for coding~\cite{chatdev,metagpt}, search~\cite{autogen,camel,li2026papers}, and simulation~\cite{generativeagents,agentsociety,virtualhome,opencity}.
Social simulation pushes multi-agent serving to its limits: it requires many agents running simultaneously, complex multi-round interactions where each agent's context grows over time, and tool use that further extends prompt length.
As these applications grow from a handful of agents to tens or hundreds in production deployments, GPU memory becomes the dominant scaling constraint. Each agent maintains a KV Cache that persists across interaction rounds, and all active agents' caches must coexist in GPU memory simultaneously. The central challenge for multi-agent LLM serving is therefore maximizing the number of agents that can run concurrently under a fixed memory budget.

\begin{figure}[t]
\centering
\includegraphics[width=\columnwidth]{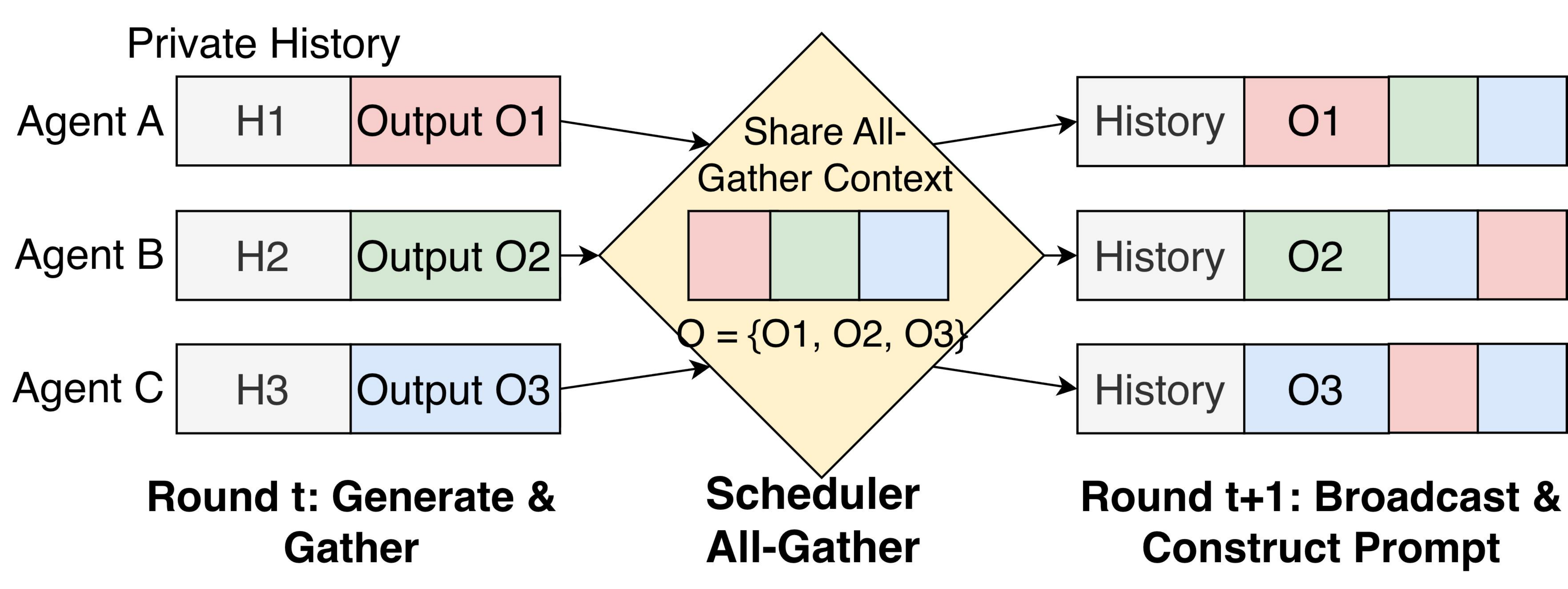}
\caption{
The \AG prompt structure.
All agents receive the same output blocks ($O$), but the blocks appear at different positions because each prompt has its own private history ($H$) and may use a different block order.
This structure arises in any multi-agent application that follows the \AG pattern.
}
\label{fig:ag-example}
\end{figure}

This memory pressure has a structural root cause.
Multi-agent frameworks such as OpenClaw ~\cite{openclaw} and MoltBook~\cite{moltbook} organize execution in synchronized rounds: each agent produces an output, a central scheduler gathers all outputs, and redistributes the combined context to every agent for the next round.
We call this recurring data flow the \textbf{\AG} pattern, after the corresponding collective in distributed computing.
The pattern directly explains the memory problem.
As Figure~\ref{fig:ag-example} shows, each agent's prompt for round $t{+}1$ combines a private history ($H$) with the full set of shared output blocks ($O$).
Because private histories differ in length, the same shared blocks land at different absolute positions across prompts, preventing the serving engine from recognizing the overlap.
The result is $N$ near-identical KV Caches in GPU memory.
Therefore, memory grows with agent count rather than with the amount of unique context.

\begin{figure}[t]
\centering
\begin{subfigure}{0.48\columnwidth}
\centering
\includegraphics[width=\linewidth]{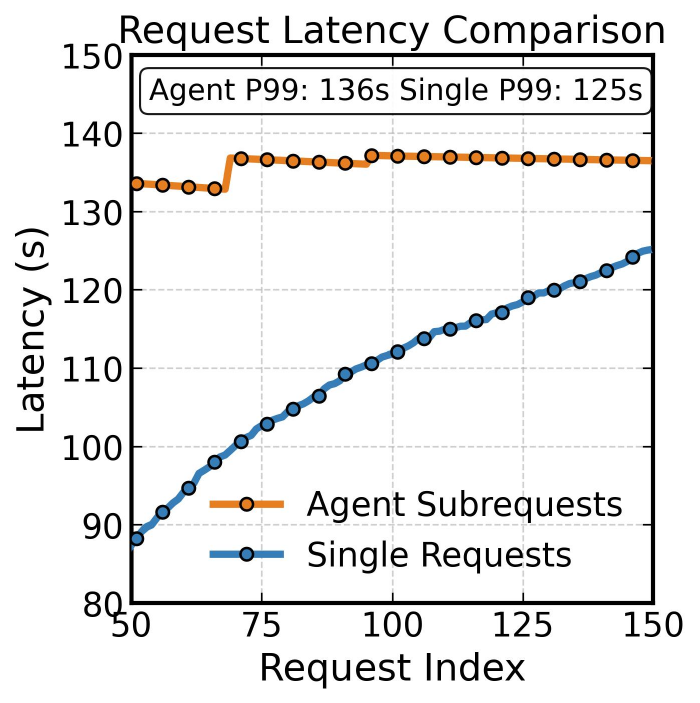}
\caption{
Subrequest latency against request index.
}
\end{subfigure}
\hfill
\begin{subfigure}{0.49\columnwidth}
\centering
\includegraphics[width=\linewidth]{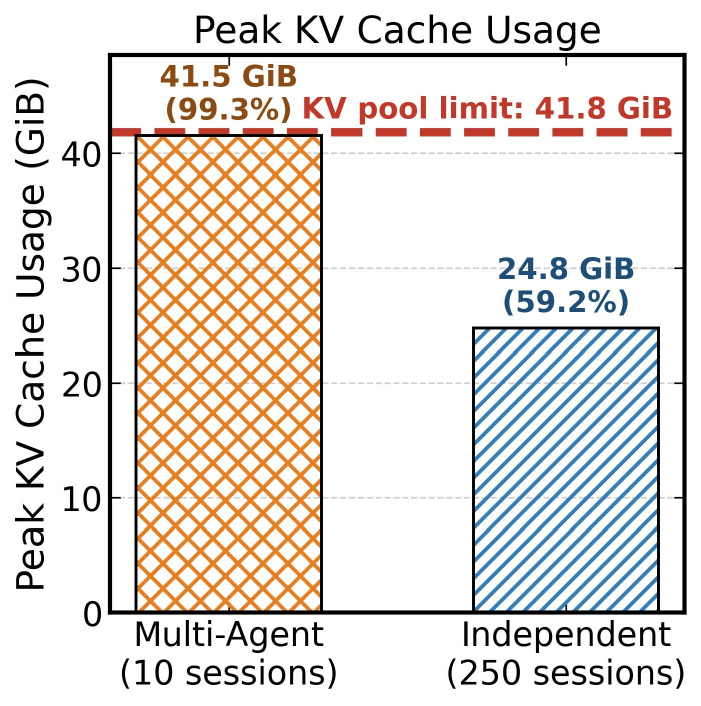}
\caption{
Peak \KVC usage for the two workloads.
}
\end{subfigure}
\caption{
The scaling gap between multi-agent and independent workloads on a single A100-80GB GPU serving Qwen2.5-14B.
Both workloads issue the same total number of subrequests (250), but the multi-agent workload nearly exhausts the \KVC pool because each agent retains its own copy of the shared context across rounds, whereas independent requests free memory after completion.
}
\label{fig:insight_capacity}
\end{figure}

Figure~\ref{fig:insight_capacity} quantifies the scale of this redundancy.
On an 80 GB A100 GPU serving Qwen2.5-14B, we compare 10 multi-agent sessions (250 agent subrequests total) against 250 independent single requests.
The multi-agent workload consumes 41.5\,GiB of \KVC storage---99.3\% of the pool---while independent requests use only 24.8\,GiB (59.2\%).
The latency impact is immediate: multi-agent sessions hit a P99 of 136 s because the memory pool saturates from the start, whereas independent requests start lower and rise gradually to 125 s.
The difference is that independent requests free their \KVCs after completion, whereas multi-agent \KVCs must coexist across rounds, saturating the pool and forcing the scheduler to preempt and swap.

Existing serving systems do not exploit this redundancy because they operate at the wrong granularity.
Systems like vLLM~\cite{vllm} and SGLang~\cite{sglang} treat each agent subrequest as an independent inference request, applying prefix caching or position-independent caching per request.
Agent-aware schedulers such as Parrot~\cite{parrot}, Autellix~\cite{autellix}, and Tokencake~\cite{tokencake} optimize \emph{when} each subrequest runs and manage offloading, but do not change \emph{how} the resulting \KVCs are computed or stored; every agent still holds a full, independent cache copy.
The missing capability is round-level \KVC optimization: detecting that agents in the same round share most of their context and using this overlap to reduce both compute and memory cost.
Exploiting this overlap, however, requires solving two specific problems.

The first problem is \emph{inefficient reuse}.
Each agent's prompt in a new round contains a private history and a shared set of output blocks from the previous round.
Because histories differ in length and the scheduler may choose different block orders, the same shared blocks appear at different absolute positions across requests.
As Figure~\ref{fig:ag-example} shows, block $O_1$ starts at a different offset in each agent's prompt.
Prefix caching~\cite{sglang, vllm} only matches tokens from position zero, so it cannot detect any sharing once private histories diverge.
Position-independent caching (PIC) methods~\cite{cacheblend,epic,kvlink,kvcomm} remove this constraint and can reuse blocks at arbitrary offsets, but they still process each request separately: in an $N$-agent round the same shared blocks trigger $N$ separate reuse passes, paying per-request overhead for what is fundamentally a round-level sharing opportunity.
Section~\ref{sec:limitations} and Figure~\ref{fig:pic-vs-collective} contrast this per-request approach with collective reuse.

\begin{figure}[t]
\centering
\begin{subfigure}{0.43\columnwidth}
  \centering
  \includegraphics[width=\linewidth]{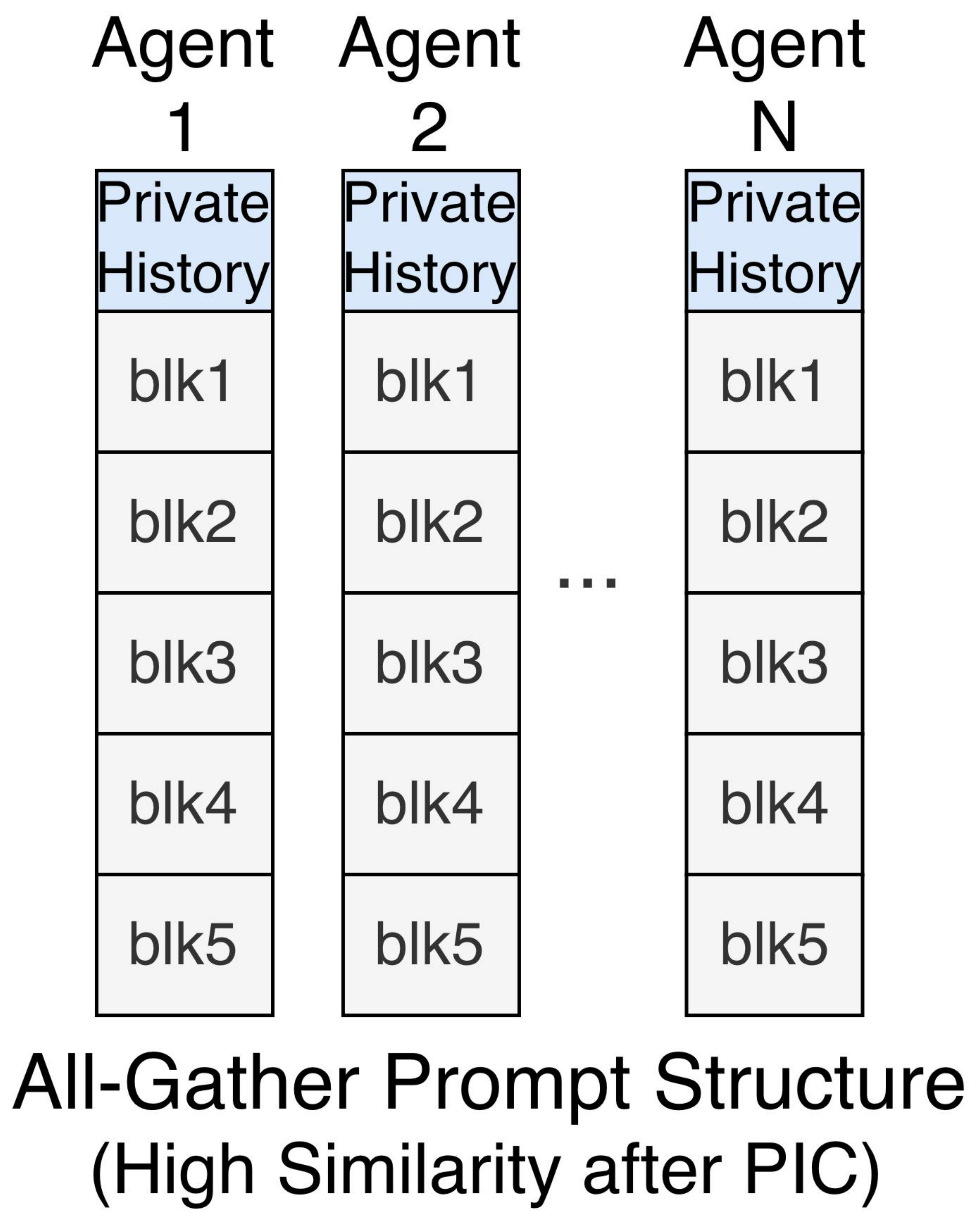}
  \caption{Schematic: Structural Overlap}
\end{subfigure}
\hfill
\begin{subfigure}{0.55\columnwidth}
  \centering
  \includegraphics[width=\linewidth]{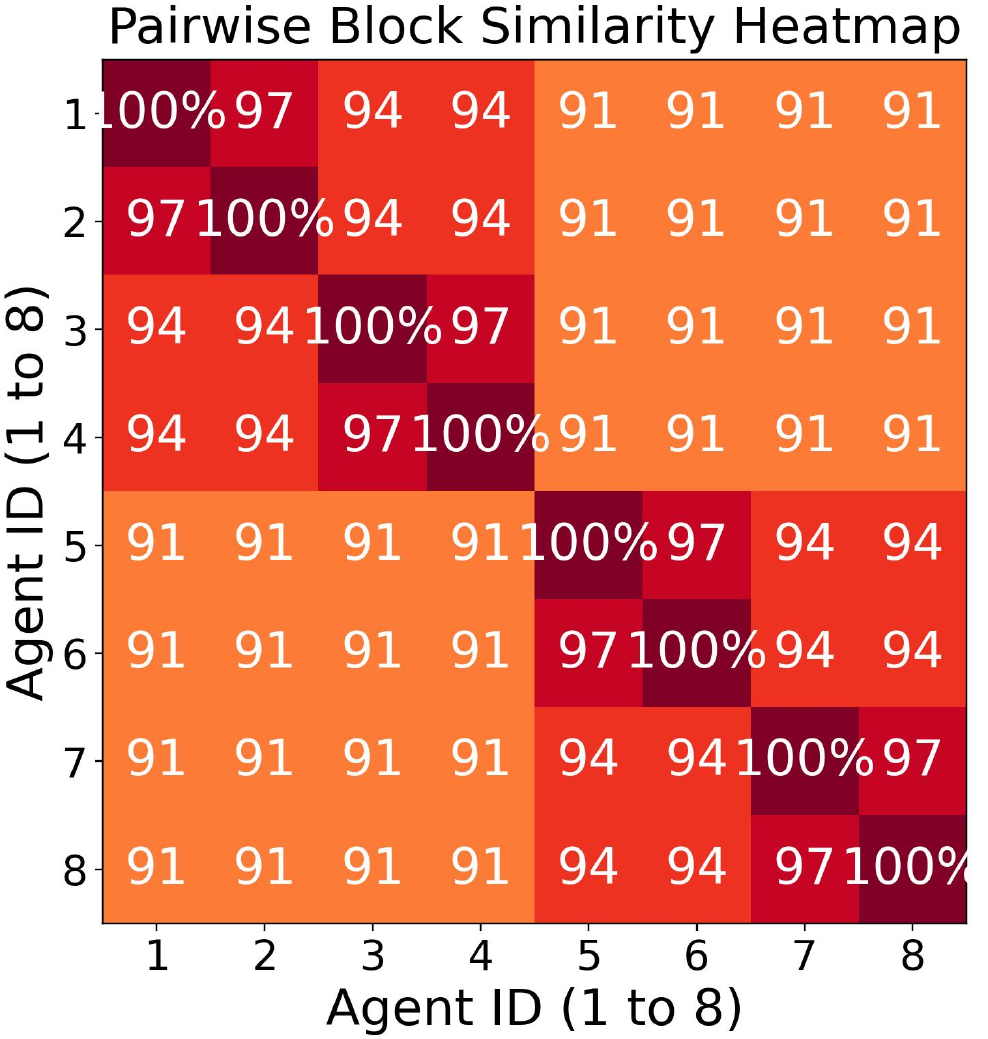}
  \caption{Measurement: Pairwise Block Similarity}
\end{subfigure}
\caption{
High similarity of \KVCs after PIC reuse.
Because all agents reuse the same shared blocks, their \KVCs differ only at the privately-recomputed positions.
}
\label{fig:insight_similarity}
\end{figure}

The second problem is \emph{inefficient storage}.
In \AG workloads where agents share the majority of their round context, the resulting \KVCs after reuse are nearly identical.
The reason is structural: the shared output blocks are reused from the same cached source, so they produce identical \KVC values across all agents; the only entries that differ are tokens that each agent recomputes from its own history, which typically account for a small fraction of the total sequence.
Figure~\ref{fig:insight_similarity} confirms this quantitatively: in an 8-agent \GA~\cite{generativeagents} round, pairwise block similarity ranges from 91\% to 97\%, meaning the vast majority of cached data is duplicated across agents.
Storing one full \KVC per agent therefore wastes most of the GPU memory budget and directly caps how many agents can remain active.
Making one round faster does not solve the problem if each agent still holds a full \KVC copy.
The per-agent storage cost must also shrink so the same hardware can sustain more concurrent agents.

We present \project, a system that scales the number of concurrent agents on local GPUs by exploiting the \AG pattern for collective \KVC sharing.
Instead of optimizing individual requests, \project operates on the multi-agent round as a whole.
It amortizes reuse computation across all agents in a round so the cost of sharing a block is paid once regardless of agent count, and it compresses \KVC storage to only the inter-agent differences so memory cost grows with the actual difference rather than with the number of agents.

We evaluate \project on the multi-agent social simulation frameworks \GA~\cite{generativeagents} and \AS~\cite{agentsociety}.
Compared to vLLM with prefix caching, \project reduces end-to-end latency by up to $2.3\times$ and \KVC storage by $94\%$.
Against CacheBlend, \project achieves a $1.9\times$ speedup in the prefill phase.
These gains translate directly into higher agent capacity: under the same latency target, \project supports up to $2.7\times$ more concurrent agents.

\section{Background}
\label{sec:background}

\subsection{The \AG Pattern}
\label{sec:ag-pattern}

As introduced in Section~\ref{sec:intro}, the \AG pattern organizes multi-agent execution in synchronized rounds where a scheduler gathers all agent outputs and redistributes the combined context.
We now formalize this structure.
For a system with $N$ agents, each agent $i$ maintains a private history $H_i^t$ containing its system prompt and prior interactions up to round $t$.
In round $t$, each agent $j$ produces an output block $O_j^{t}$.
The shared set of round outputs is:
\begin{equation}
\label{eq:round_context}
\mathcal{O}^{t} = \{O_1^{t}, O_2^{t}, \ldots, O_N^{t}\}
\end{equation}
The prompt for agent $i$ in round $t+1$ is then:
\begin{equation}
\label{eq:prompt_composition}
P_i^{t+1} = H_i^t \Vert \Pi_i(\mathcal{O}^{t})
\end{equation}
where $\Pi_i$ is the scheduler-defined layout of the shared output blocks for agent $i$.
Every prompt in the round contains the same shared output blocks and one private history, and the shared blocks appear at different absolute positions across requests because the private histories differ in length.

\subsection{\KVC Reuse in LLM Serving}
\label{sec:llm-inference}

LLM serving is memory-bound: the system stores \KVCs to avoid recomputing attention keys and values for prior tokens, and managing these caches is the primary challenge for scaling system capacity.
Existing reuse methods fall into two categories.

Prefix caching, used by vLLM~\cite{vllm} and SGLang~\cite{sglang}, reuses cached tensors only when a new request shares an exact token prefix with a stored sequence.
This works well for requests that share a common system prompt but fails once per-agent private histories cause the prefix to diverge.

Position-independent caching (PIC) methods~\cite{cacheblend,epic,kvcomm,kvlink} remove this constraint by recovering shared blocks at arbitrary offsets.
The key idea is to correct for positional encoding differences: a PIC method first applies RoPE rotation to align cached Keys with the target positions in the new request, then computes key differences between the rotated cached values and freshly computed values to identify \emph{important positions}---token positions where the two differ significantly.
Only these important positions are selectively recomputed, while the remaining positions reuse the rotated cached values directly.
This recovers sharing beyond the prefix at the cost of additional compute for rotation and recomputation, and a minor accuracy trade-off at positions that are reused without recomputation.

\begin{figure}[t]
  \centering
  \includegraphics[width=\columnwidth]{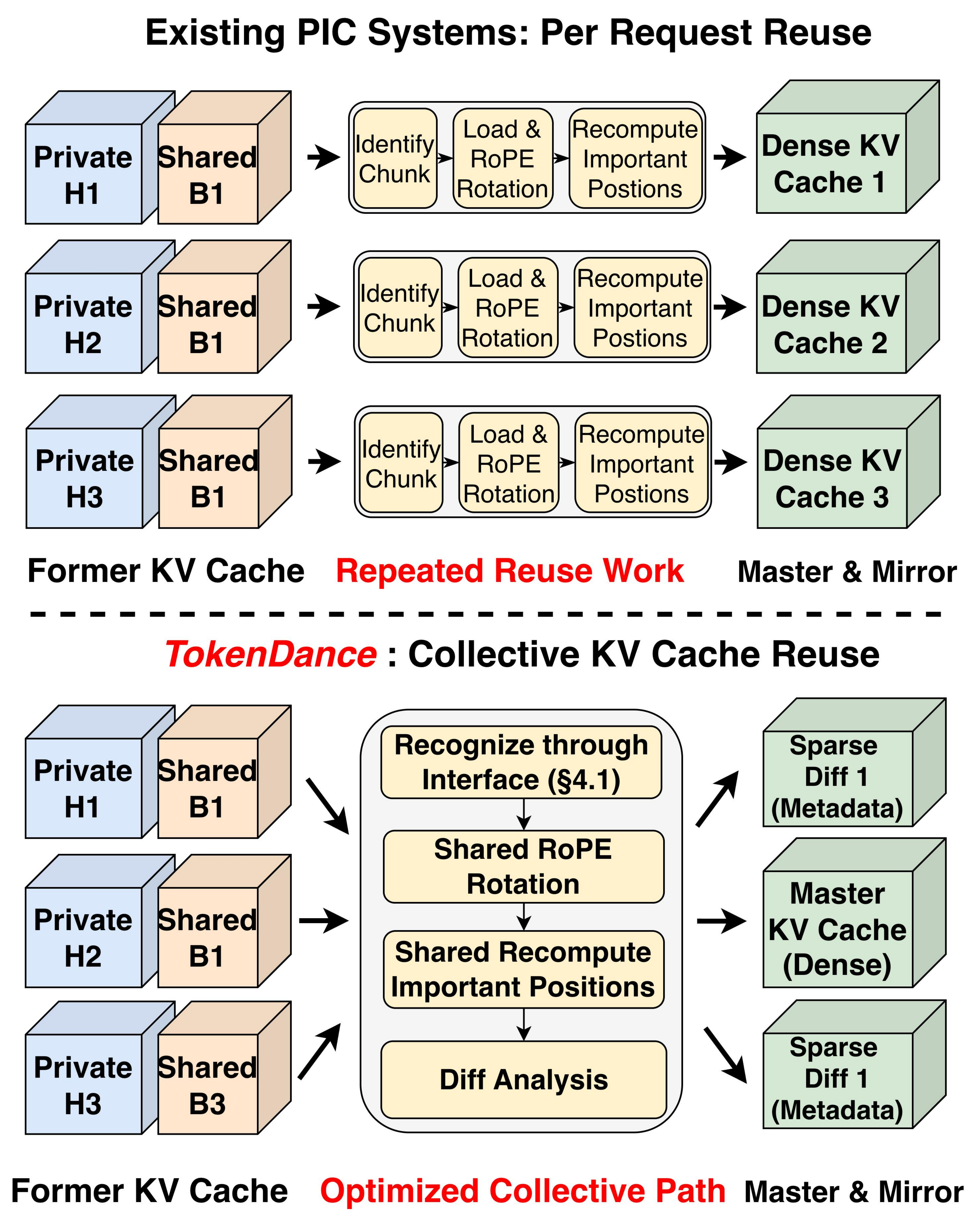}
  \caption{Per-request PIC reuse (top) vs.\ \project's collective reuse (bottom).
Existing PIC methods process each agent's shared blocks independently, repeating RoPE rotation and important-position selection $N$ times.
\project groups the $N$ requests and performs these operations once for the round.}
\label{fig:pic-vs-collective}
\end{figure}

\subsection{Limitations of Existing Serving Systems}
\label{sec:limitations}

The \AG pattern creates a specific combination of redundancy that existing serving systems fail to exploit.
We group prior work into two categories and explain why each falls short.

\noindent\textbf{Agent-Aware, Compute-Centric Schedulers.}
Parrot~\cite{parrot}, Autellix~\cite{autellix}, and Teola~\cite{teola} inject application-level dependency graphs into the serving backend to prioritize critical requests and reduce head-of-line blocking.
ScaleSim~\cite{scalesim} estimates future invocation distances to guide prefetching in simulation workloads.
These systems decide \emph{when} each request runs, but they do not control \emph{how} the resulting \KVCs are computed or stored.
The memory allocator underneath still treats every agent's cache as an independent object.
A burst of non-critical agents can therefore evict a critical agent's cache regardless of the schedule.
None of these systems detects that the $N$ requests in an \AG round share the same output blocks, so neither the reuse work nor the storage footprint is reduced.

\noindent\textbf{\KVCc, Agent-Agnostic Engines.}
vLLM~\cite{vllm} and SGLang~\cite{sglang} reuse cached tensors through prefix matching, which breaks as soon as private histories diverge from position zero.
CacheBlend~\cite{cacheblend} and EPIC~\cite{epic} lift this positional constraint by reusing \KVC blocks at arbitrary offsets, but they still analyze each request separately: $N$ agents trigger $N$ independent reuse passes over the same shared blocks.
Mooncake~\cite{mooncake} and CacheGen~\cite{cachegen} expand the cache pool through disaggregation and compression, yet their policies are driven by global memory pressure, not by round structure.
Tokencake~\cite{tokencake} adds agent-aware scheduling on top of a \KVCc backend, proactively offloading stalled agents and reserving memory for critical ones.
Across all of these systems, the final outcome is the same: every agent holds a dense, independent \KVC copy, and no storage is freed even when over 90\% of cache blocks are identical across agents in the same round.

Figure~\ref{fig:pic-vs-collective} contrasts this per-request approach (top) with the collective alternative that \project introduces (bottom).

\section{Overview}
\label{sec:overview}

\begin{figure}[t]
\centering
\includegraphics[width=\columnwidth]{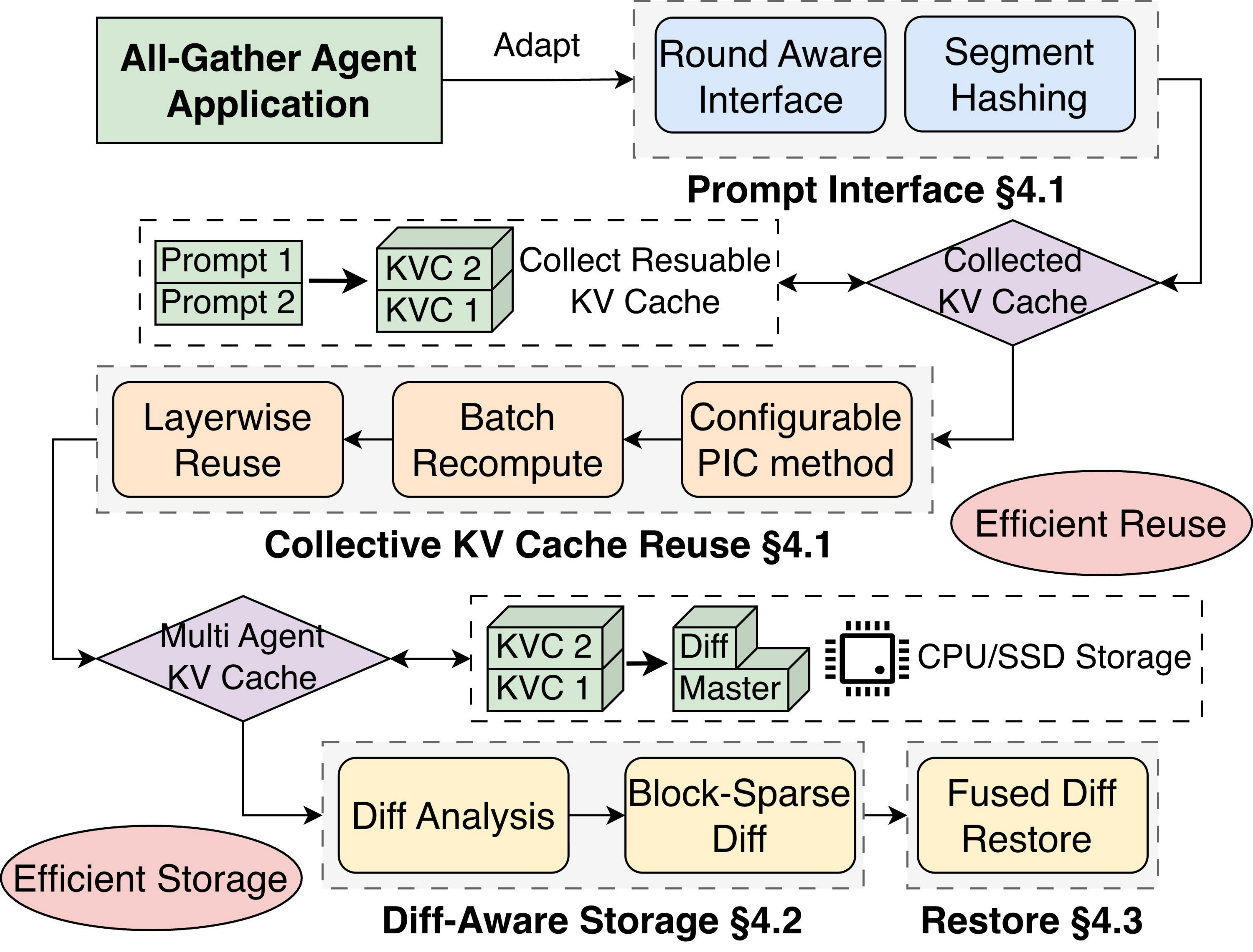}
\caption{
\project Overview.
A round-aware prompt interface preserves block boundaries so the runtime can identify shared content; collective \KVC reuse amortizes the reuse cost across all agents in the round; diff-aware storage with fused restore compresses per-agent \KVCs to only the inter-agent differences.
}
\label{fig:overview}
\end{figure}

\project scales the number of active agents that a local GPU can support in an \AG system by addressing both redundant reuse computation and redundant \KVC storage.
The key design principle is to lift the unit of optimization from a single request to the \AG round: every operation that is common across agents in a round is performed once and shared, while only the per-agent differences are handled individually.
Figure~\ref{fig:overview} illustrates the three components that realize this principle.

The round-aware prompt interface (Section~\ref{sec:prompt_interface}) preserves the logical block structure of each prompt by inserting separator tokens between adjacent blocks, so the runtime can recognize shared blocks even when they appear at different absolute positions.
The collective \KVC reuse algorithm (Section~\ref{sec:collective_reuse}) exploits this structure to perform reuse for a compatible group of requests in one pass, paying the reuse overhead once per round instead of once per request.
The diff-aware storage scheme (Section~\ref{sec:diff_storage}) encodes sibling caches as sparse differences against a shared Master copy, and the fused restore path (Section~\ref{sec:fused_diff}) applies those differences during GPU transfer to avoid a separate reconstruction step.
Together, these components reduce both the compute work and the storage cost that each additional agent adds, allowing the same hardware to sustain more agents at the same service level.

\section{\project Design}
\label{sec:design}

The design of \project starts from a mismatch between the \AG application structure and the serving system.
Multi-agent applications communicate in rounds, but the serving system receives a flat token stream, loses the logical boundary between private and shared blocks, and treats every request as independent.

This mismatch creates three compounding inefficiencies.
First, the runtime cannot identify shared blocks across requests because position-based indexing conflates content identity with absolute offset.
Second, position-independent reuse methods repeat the same analysis for each request in the round, even though all requests consume the same shared blocks.
Third, even after reuse, the resulting \KVCs are nearly identical, yet the system stores one full dense copy per request.
The cumulative result is an inflated per-agent \KVC cost that directly caps the number of concurrent agents.

\project addresses these inefficiencies with two design rules.
First, it keeps the round structure visible until the runtime can use it.
Second, it preserves one correct cache state per request while sharing the work and state that are truly common across the round.
The first rule leads to a round-aware interface (Section~\ref{sec:prompt_interface}) and collective \KVC reuse (Section~\ref{sec:collective_reuse}).
The second rule leads to diff-aware storage and fused diff restore (Section~\ref{sec:fused_diff}).
The rest of this section describes each component in detail.

\subsection{Round-Aware Prompt Interface}
\label{sec:prompt_interface}

The runtime cannot exploit the \AG pattern if it only receives a flat token stream.
\project provides a round-aware prompt interface that preserves the logical block structure of each prompt, so the runtime can recognize shared blocks even when they appear at different absolute positions across requests.

Any multi-agent framework that follows the \AG pattern can adopt this interface with minimal code changes.
If the application does not follow the pattern, \project falls back to the standard single-request path with no performance loss.

\begin{figure}[t]
\centering
\includegraphics[width=\columnwidth]{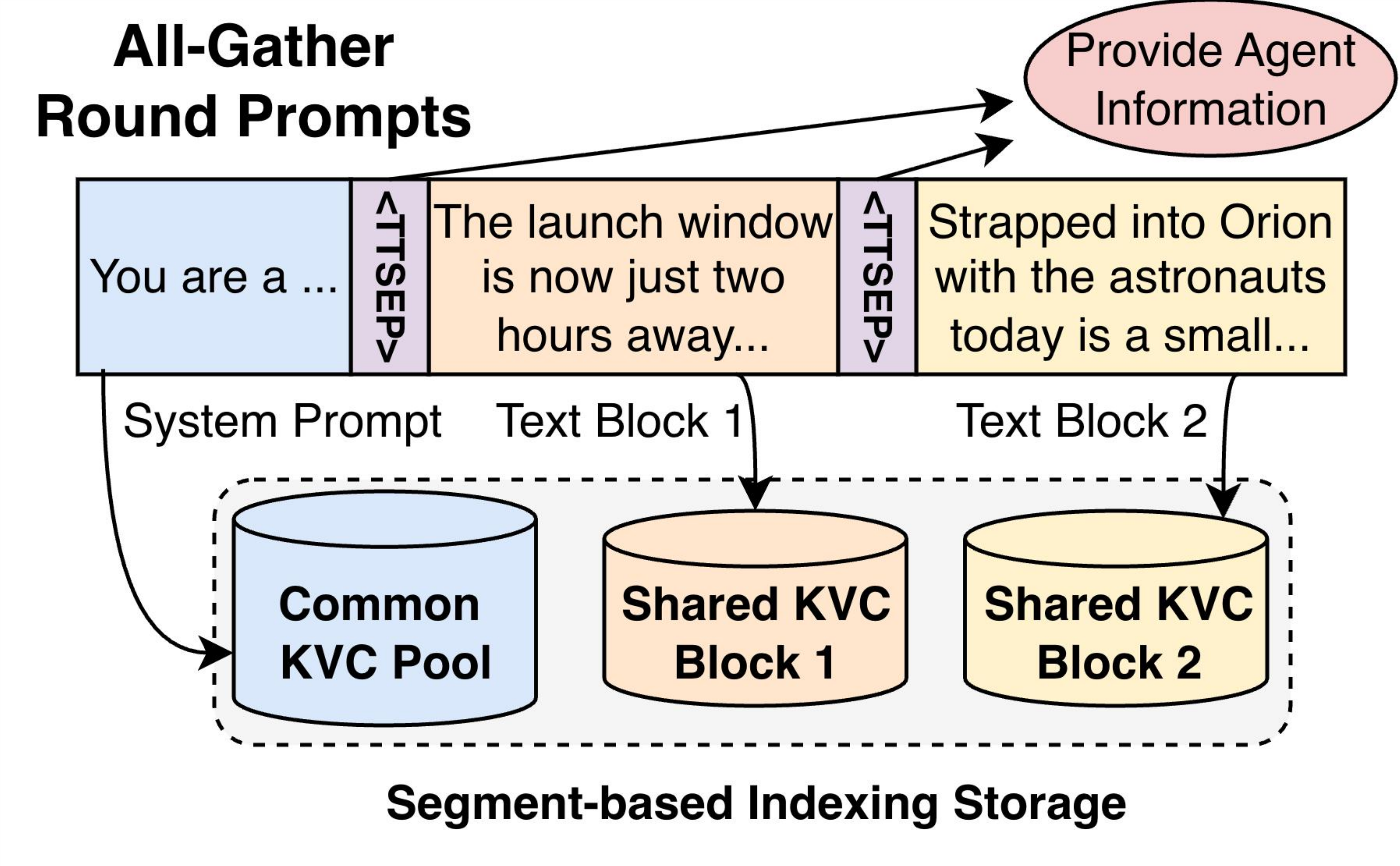}
\caption{
Example of \project's round-aware prompt interface.
Each prompt is composed of a private history block and a shared set of output blocks, with reserved separator tokens (\texttt{<TTSEP>}) in between.
}
\label{fig:prompt_interface}
\end{figure}

The application assembles one prompt per agent and inserts a reserved separator token \texttt{<TTSEP>} between adjacent logical blocks.
Figure~\ref{fig:prompt_interface} shows an example for three agents in one round.
Each prompt contains a private history block and the same set of shared output blocks, with separator tokens in between.
The delimited blocks make the reusable relationship between requests explicit: the private history, each shared agent update, and the round task are all individually identifiable after tokenization.

Once block boundaries are visible, the runtime switches from fixed-size chunk hashing to segment-based hashing.
Each shared update is indexed by its own content segment rather than by its absolute position in one long prompt.
Two requests that contain the same shared update therefore map that update to the same cache object, even when their private histories differ in length.
This is the first step that turns the \AG pattern into a serving optimization: the runtime now knows which blocks are shared and can reason about reuse at the block level rather than the prefix level.

\subsection{Collective \KVC Reuse}
\label{sec:collective_reuse}

Once the round structure is visible, the next problem is repeated reuse work.
Existing position-independent caching (PIC) methods~\cite{cacheblend,epic,kvlink,kvcomm} can recover sharing beyond the prefix, but they still process each request independently.
In an $N$-agent round, the runtime performs $N$ separate reuse passes for the same set of shared blocks.
Each pass applies RoPE, computes key differences, and selects important positions---all operations that produce nearly identical results across requests because the shared blocks are content-identical.
This per-request redundancy is a poor match for the \AG pattern, where the shared round updates are common to every request in the round.

\begin{figure}[t]
\centering
\includegraphics[width=\columnwidth]{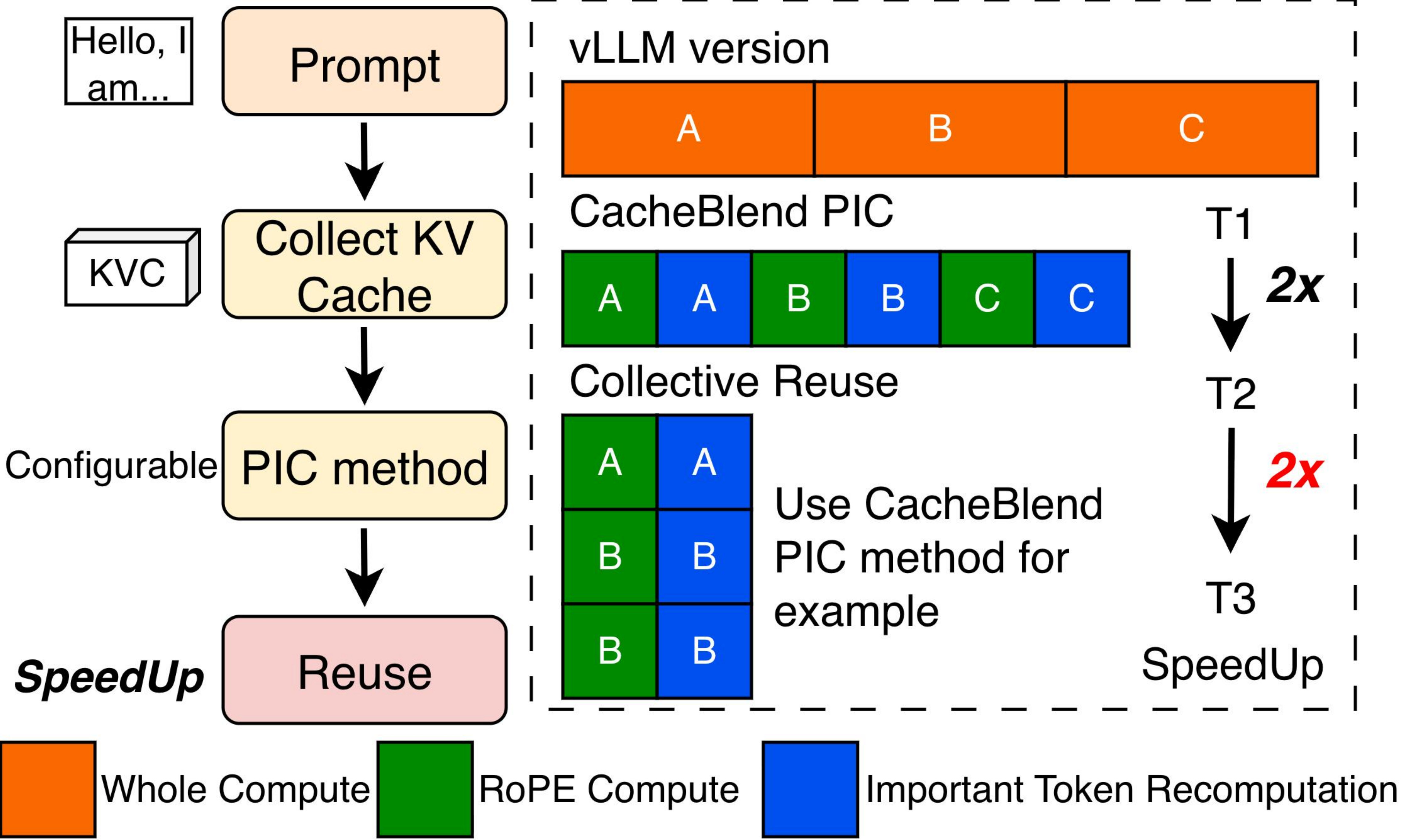}
\caption{
Collective \KVC reuse for a three-agent \AG round.
Left:
each agent's prompt contains a private section and the same shared blocks in different orders.
Right:
vLLM computes all three from scratch (T1);
per-request PIC processes each request independently (T2);
\project groups them and shares the RoPE and important-position selection work across the group (T3), paying the reuse overhead once per round.
}
\label{fig:kv_collector}
\end{figure}

\project resolves this redundancy with collective \KVC reuse, illustrated in Figure~\ref{fig:kv_collector}.
Rather than processing each request independently, the \KVCollector groups the $N$ requests in a round and performs one shared RoPE rotation and one shared important-position selection pass, then updates only the differing positions in each request's cache.
Figure~\ref{fig:kv_collector} contrasts this collective path (T3) with full recomputation (T1) and per-request PIC reuse (T2).

\noindent\textbf{Grouping Compatible Requests.}
The \KVCollector collects requests from the same \AG round whose prompt spans are compatible for collective processing.
Compatibility requires that the requests have the same active prompt length, the same cached span visible to the cache layer, and non-overlapping slot mappings in the execution engine.
These are execution constraints that ensure lockstep layerwise processing; requests that do not satisfy them fall back to separate groups or to the single-request path.

\noindent\textbf{Layerwise Collective Reuse.}
\label{sec:layer_collective}
For each compatible group, the runtime drives layerwise retrieval and model execution in lockstep.
At each layer, \KVCollector concatenates the Q and K tensors from all requests in the group into one combined tensor and applies a single batched RoPE call.
On the configured check layer, the runtime compares the rotated Keys against the cached Keys for the entire group in one batched difference pass.
This pass identifies the important positions---the token positions where the cached values diverge significantly from the freshly computed values---for each request simultaneously.
The runtime then refreshes only those important positions in each request's cached K and V tensors.
Later layers reuse the important-position set identified on the check layer directly, without repeating the difference computation.

The expensive operations---RoPE rotation and key-difference analysis---are therefore performed once for the group rather than once per request.
Only the final cache update remains request-specific, because each request's private history produces different values at the positions that need refreshing.
In a request-centric path, the reuse-analysis overhead grows linearly with the number of agents.
In \project, the group pays one shared RoPE pass and one shared diff-analysis pass per layer, and only the per-position refresh scales with agent count.
As the number of agents in a round grows, this amortization reduces the total reuse work per round.
That reduction is the compute side of the scaling gain.

The collective amortization is decoupled from the underlying per-position recovery method.
The current prototype uses CacheBlend~\cite{cacheblend}'s selective recomputation as the default backend, but any PIC method that accepts a set of token positions and returns corrected K/V tensors can serve as a drop-in replacement through an adapter interface.

\noindent\textbf{Reuse Plan Output.}
\label{sec:reuse_plan}
Collective \KVC reuse also produces metadata that \DStorer later consumes.
The reuse path records the group membership, the accumulated per-request deviation scores, and the request chosen as the Master.
The Master is not semantically special.
It is simply the request whose recovered result is closest to the group's common structure, typically the one with the lowest total deviation from the shared blocks.
Selecting a good Master minimizes the size of the sparse corrections that \DStorer must keep for the remaining requests.
The reuse plan therefore serves as the bridge between the compute optimization in this subsection and the storage optimization in the next.

\subsection{Diff-Aware Storage}
\label{sec:diff_storage}

Collective \KVC reuse removes redundant compute, but it does not remove redundant storage.
After reuse completes for one round, the system still holds one dense \KVC per request.
This is wasteful because the requests in one \AG round share most of their logical blocks and differ only in a private history and a small number of position-sensitive updates.
If the system stores every recovered result in full, it merely shifts redundancy from prefill compute to memory capacity.

This problem is especially acute for multi-agent serving because rounds repeat and agent states stay live across rounds.
In a system with $N$ agents running over $R$ rounds, storing one full cache per agent per round leads to memory consumption that grows as $O(N \times R)$.
If the system stores many near-duplicate \KVCs, memory pressure rises quickly and the number of active agents falls.
Storage must therefore exploit the same \AG structure that reuse exploits: the requests in a round are structurally similar and differ only at a predictable subset of positions.

\begin{figure}[t]
\centering
\includegraphics[width=\columnwidth]{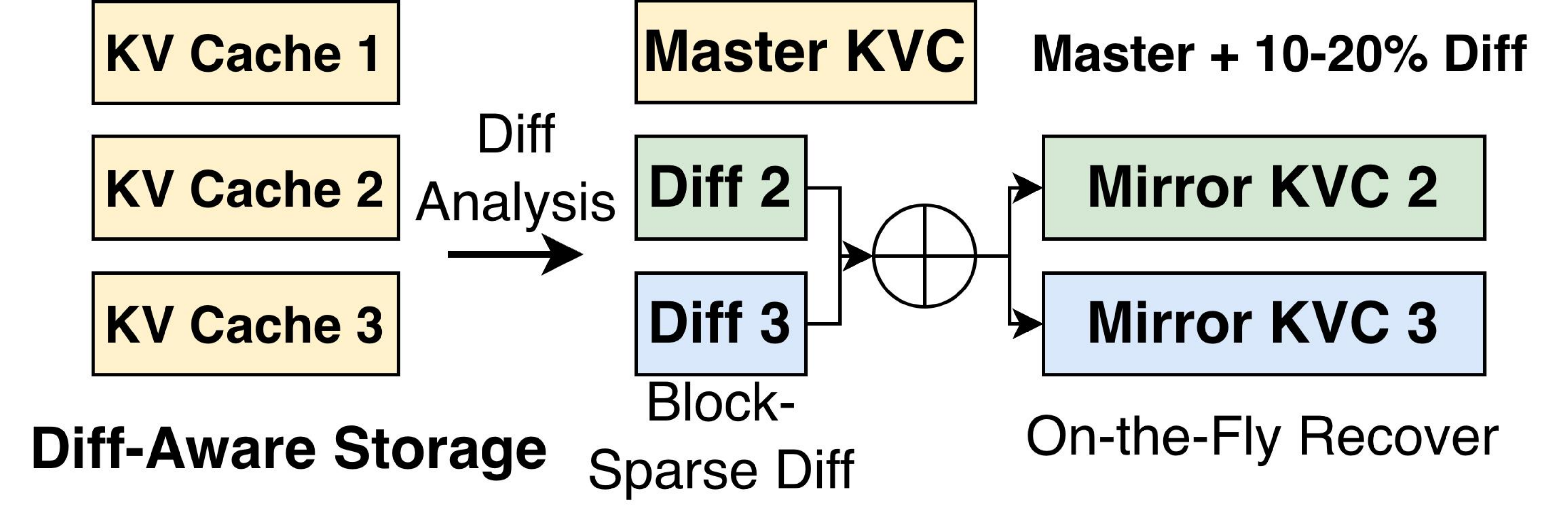}
\caption{
Diff-aware storage with the Master-Mirror layout.
Left: after reuse, the \KVCs of three agents differ at only 10--20\% of positions.
Right: \DStorer stores one full Master cache and encodes each remaining cache as a sparse diff (Diff 2, Diff 3).
On restore, the system reconstructs each Mirror on the fly from the Master plus its diff.
}
\label{fig:block_storage}
\end{figure}

\noindent\textbf{Master-Mirror Layout.}
\label{sec:master_mirror_layout}
\project uses a Master-Mirror layout to compress the per-round cache family, as shown in Figure~\ref{fig:block_storage}.
\DStorer keeps one dense cache as the Master and stores each remaining cache as a Mirror that contains only the difference from that Master.
The adapter forwards the reuse plan from collective \KVC reuse into the store path, so storage already knows which request should become the Master and which positions differ in each Mirror.
When no explicit reuse plan is available---for example, when a request arrives outside a recognized \AG round---the backend falls back to a token-similarity heuristic to find a reusable Master among existing cached entries.

\noindent\textbf{Block-Sparse Diff Representation.}
\label{sec:block_diff_representation}
The Mirror correction is stored as a block-sparse K/V diff.
The differences among recovered caches are concentrated in a subset of token positions whose values changed because of private context or position-dependent RoPE rotation.
These differing positions tend to cluster in contiguous blocks that correspond to the private history segment or to the boundary regions between shared blocks.
A block-granular representation captures this clustering with much lower metadata cost than a fine-grained element-wise diff.
It also aligns naturally with the block-based memory management used by modern serving engines and with the tile-aligned restore pipeline described in Section~\ref{sec:fused_diff}.

On read, the storage layer does not eagerly rebuild a dense Mirror tensor.
Instead, it returns a lightweight mirror object that keeps a reference to the Master and the sparse diff metadata.
This preserves a standard logical \KVC abstraction for the caller while delaying materialization until the runtime actually needs the data.
The Mirror occupies only 10--20\% of a full cache, so the system saves memory immediately without paying the restore cost until execution time.

The strength of this layout depends on how much of the prompt is shared versus private.
The \AG pattern is a natural fit because the shared output blocks from the previous round dominate the prompt, and only the private history and round task are unique per agent.
If requests diverge more strongly---for example, in a system where each agent receives a substantially different subset of the shared outputs---then the Mirror corrections grow larger and the storage benefit diminishes.
We quantify this relationship in Section~\ref{sec:eval-memory}.

\subsection{Fused Diff Restore}
\label{sec:fused_diff}

Compression is useful only if restore stays cheap on the critical path.
A naive Mirror restore would load the full Master, copy it into a new buffer, and overwrite the differing blocks---adding an extra dense write-then-read round trip for an object the system never keeps.

\project eliminates this overhead by applying the sparse corrections inside the layerwise transfer path that already moves cached KV data into paged GPU memory.
Algorithm~\ref{alg:fused_diff} shows the procedure.
Two GPU buffers alternate roles in a ping-pong fashion: one receives Master chunks from storage while the other undergoes in-place correction and writeback.
At each layer, the path applies the block-sparse diff (line~\ref{line:apply_diff}), recovers RoPE positions (line~\ref{line:rope}), and writes the result into paged \KVC memory (line~\ref{line:wb})---all within the same pass that ordinary cache loads already perform.
No separate dense Mirror is ever materialized.

\begin{algorithm}[t]
\caption{Fused Diff Restore for One Mirror Request}
\label{alg:fused_diff}
\begin{algorithmic}[1]
\Require Mirror $M$, slot map $S$, positions $P^{\mathit{old}}, P^{\mathit{new}}$, layers $L$.
\State Allocate ping-pong buffers $B^{\mathit{load}}, B^{\mathit{comp}}$.
\For{$\ell \gets 1$ \textbf{to} $L$}
    \State Load Master chunks of layer $\ell$ into $B^{\mathit{load}}$.
    \State Sync; swap $B^{\mathit{load}} \leftrightarrow B^{\mathit{comp}}$.
    \State $D \gets \Call{GetDiff}{M, \ell}$
    \If{$D \neq \varnothing$}
        \State Update $B^{\mathit{comp}}.\{K,V\}$ at $D.\mathit{indices}$ with $D.\mathit{values}$. \label{line:apply_diff}
    \EndIf
    \State $B^{\mathit{comp}}.K \gets \Call{RoPERecover}{P^{\mathit{old}}, P^{\mathit{new}}, B^{\mathit{comp}}.K}$ \label{line:rope}
    \State Write $B^{\mathit{comp}}$ to paged \KVC at layer $\ell$ using $S$. \label{line:wb}
\EndFor
\end{algorithmic}
\end{algorithm}

\begin{figure}[t]
\centering
\includegraphics[width=\columnwidth]{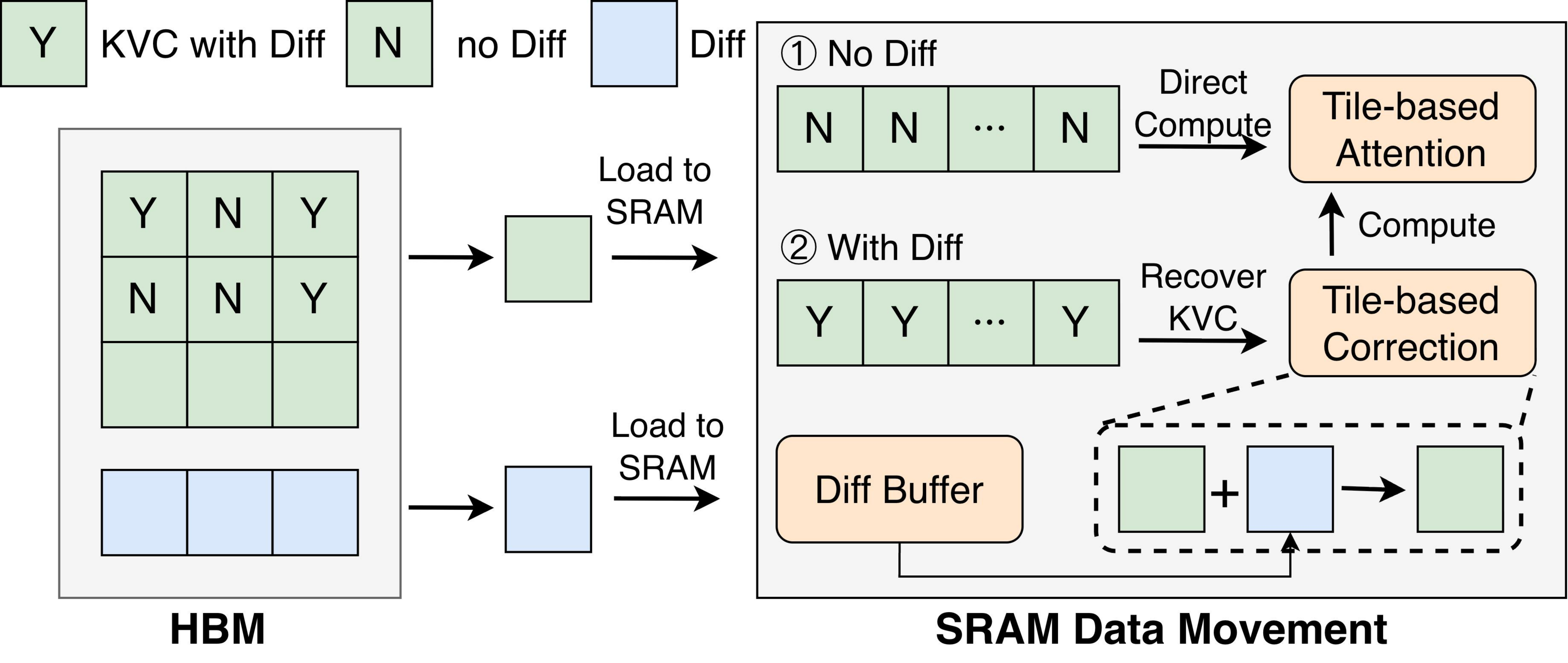}
\caption{
Fused diff restore at block granularity.
Blocks marked ``Have Diff'' are corrected in SM memory before attention; blocks with no diff bypass the correction path.
}
\label{fig:sparse_kernel}
\end{figure}

The only additional work per layer is the correction at line~\ref{line:apply_diff}, whose cost is proportional to the number of differing blocks---typically 10--20\% of the total.
Figure~\ref{fig:sparse_kernel} illustrates the block-level dispatch.
Blocks identical to the Master proceed directly to attention.
Blocks that carry diffs are corrected in SM memory before being fed to FlashAttention.
The block-sparse format makes this skip-or-correct decision per block without scanning the full cache, and the block size aligns with attention tile sizes so corrected blocks require no additional reshaping.

The current prototype fuses the diff into the transfer pipeline before attention, not inside the attention tile loader.
This already removes the dense reconstruction step and reduces extra memory traffic on the critical path.
A deeper fusion---applying diffs as attention tiles are loaded from HBM into shared memory---is a natural extension.

Taken together, diff-aware storage and fused restore address the memory side of the scaling problem.
Storage reduces how much state each additional agent contributes to the memory footprint.
Fused restore ensures this compression remains usable during online execution rather than only at rest.
This is how \project turns the structural similarity inherent in the \AG pattern into higher active-agent capacity on local GPUs.

\section{Implementation}
\label{sec:implementation}

\project is implemented on top of LMCache and vLLM.
We add approximately 3K lines of Python and 500 lines of CUDA/C++ across four parts: round-aware segment indexing, collective \KVC reuse, diff-aware storage, and fused sparse restore.

\noindent\textbf{Round-Aware Segment Indexing.}
The application inserts \texttt{<TTSEP>} separators between logical blocks as described in Section~\ref{sec:prompt_interface}.
On the runtime side, we replace the fixed-size chunk hash table with a segment-based hash table that splits prompts at separator boundaries and indexes each segment independently.

\noindent\textbf{Collective \KVC Reuse.}
The vLLM V1 adapter scans requests in each scheduler step, and dispatches each group to the collective path described in Section~\ref{sec:collective_reuse}.
Internally, the \KVCollector creates one compute generator and one retrieval generator per request and drives them in lockstep across layers.

\noindent\textbf{Diff-Aware Storage.}
We add a diff-aware backend that wraps the normal LMCache storage backend.
When a reuse plan is available, the store path uses it to select the Master and identify Mirror positions directly.
When no reuse plan is available, the backend falls back to a token-similarity heuristic.
Master chunks are written unchanged.
Each Mirror chunk is serialized as a block-sparse K/V diff: the backend records the touched block indices and the K/V correction values for those blocks.
When both K and V planes touch the same blocks, the implementation shares the block-index list between them to reduce metadata size.
On read, the backend returns a lightweight mirror object rather than a dense tensor, deferring materialization to the restore path.

\noindent\textbf{Fused Sparse Restore.}
We extend the GPU connector's layerwise ping-pong transfer path to recognize mirror objects.
The connector loads Master chunks into its temporary buffer, merges the sparse diff metadata for the current layer, and invokes a paired K/V diff kernel in place before RoPE recovery and paged-memory writeback.
The CUDA extension provides two kernels: one for a single KV plane and one that updates K and V together.
The paired kernel is used when the sparse metadata is well aligned; otherwise the connector falls back to dense restore.
As discussed in Section~\ref{sec:fused_diff}, fusion currently occurs in the transfer pipeline before attention, not inside the attention tile loader.

\section{Evaluation}
\label{sec:evaluation}

The main question of our evaluation is not whether \project speeds up one round, but whether it can scale the number of active agents under the same service target.
We organize the evaluation around five questions:
(Q1)~the system-level scaling benefit (\S\ref{sec:eval-scale});
(Q2)~where this benefit comes from on the compute side (\S\ref{sec:eval-compute});
(Q3)~where it comes from on the memory side (\S\ref{sec:eval-memory});
(Q4)~whether the fused retrieval path preserves these gains on the critical path (\S\ref{sec:eval-fused});
and (Q5)~whether \project preserves model output fidelity (\S\ref{sec:eval-accuracy}).

\subsection{Setup and Metrics}
\label{sec:eval-setup}

We evaluate \project on NVIDIA A100 $80$ GB GPU.
We use two models, Qwen2.5-7B and Qwen2.5-14B, to show how gains change with model size.
We replay traces from \GA and \AS, two representative \AG-pattern frameworks that span different operating regimes: \GA uses shorter private histories and fewer agents per round, while \AS uses longer histories with more agents.

We compare against three baselines that span the space of existing reuse strategies:
(1)~vLLM with prefix caching, which represents the standard request-local reuse path;
(2)~CacheBlend without per-request \KVC recovery (\texttt{CacheBlend Ordinary Path}), which uses a CPU-side \KVC pool but no cross-prefix reuse;
and (3)~CacheBlend with full per-request \KVC recovery, the most widely adopted open-source PIC method~\cite{cacheblend}, which performs selective cross-prefix reuse without round-level sharing.

Our metrics reflect the scaling goal directly.
We derive two capacity metrics from round latency (the end-to-end time to serve one agent round), peak \KVC footprint, and restore overhead.
The first is the maximum number of agents sustained below a fixed latency target ($1500$ ms).
The second is the maximum number of agents sustained at a given QPS before latency exceeds that target.
These two views correspond to the two ways a serving system is provisioned: by latency SLO and by throughput budget.

\begin{figure*}[t!]
\centering
\includegraphics[width=\linewidth]{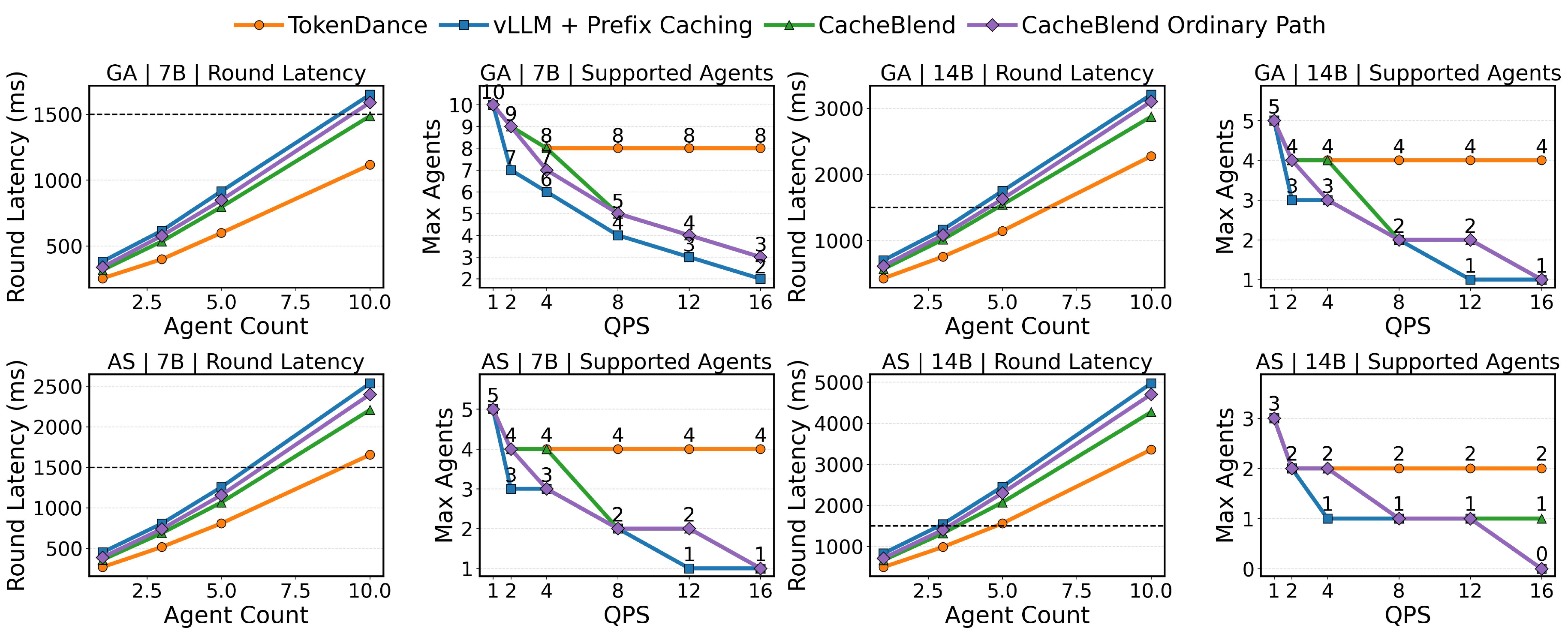}
\caption{
Scaling capacity overview across two workloads (\GA, \AS) and two models (Qwen2.5-7B, Qwen2.5-14B).
Left panels: round latency vs.\ agent count at $QPS = 10$; the dashed line marks the $1500$ ms SLO.
Right panels: maximum number of agents that stay below the SLO at each QPS level.
\project (orange) consistently supports more agents than all baselines across the full QPS range.
}
\label{fig:exp_scale}
\end{figure*}

\subsection{Main Result: Scaling the Number of Active Agents}
\label{sec:eval-scale}

Under the same latency target or QPS target, how many agents can each system support?
For each system, we sweep the number of agents from 1 to 10 and the offered QPS from 1 to 16, recording round latency at each operating point.
From these sweeps, we derive the two capacity views described above.

Figure~\ref{fig:exp_scale} presents the results across four configurations (two workloads $\times$ two models).
Each configuration shows two panels: round latency against agent count at $QPS = 10$ (left, with a dashed $1500$ ms SLO line) and maximum supported agents against QPS (right).

\project consistently achieves the lowest round latency across all configurations and agent counts.
On \GA/Qwen2.5-7B, \project keeps round latency below the SLO for all 10 agents, while vLLM with prefix caching and CacheBlend Ordinary Path cross the SLO at 10 agents.
On \GA/Qwen2.5-14B, \project still sustains all 10 agents below the SLO, while vLLM exceeds it at 7.5 agents and both CacheBlend variants exceed it at similar points.
The gap widens with model size because the per-agent \KVC footprint grows, and \project's deduplication provides proportionally larger savings.

The supported-agents panels quantify the scaling benefit.
On \GA/Qwen2.5-7B at $QPS = 16$, \project supports 8 agents while CacheBlend drops to 4 and vLLM drops to 2.
On \GA/Qwen2.5-14B, \project maintains 4 agents across the full QPS range, while vLLM drops to 1 agent beyond $QPS = 8$.
The \AS workload (bottom row) shows the same pattern: at $QPS = 4$ on \AS/Qwen2.5-7B, \project supports 4 agents while vLLM and CacheBlend Ordinary Path support 3.
On \AS/Qwen2.5-14B, the baselines can barely sustain a single agent beyond $QPS = 4$, whereas \project still supports 2 agents at $QPS = 16$.

Two trends are worth highlighting.
First, \project's advantage grows with agent count.
At small agent counts (1--2), all systems perform similarly because there is little cross-agent redundancy to exploit.
As the agent count rises, the baselines pay linearly growing reuse and storage costs, while \project amortizes both across the round.
Second, \project's advantage is more pronounced on the 14B model than on the 7B model.
This is because the per-agent \KVC footprint doubles with model size, making deduplication more valuable in absolute terms.
These results confirm the central claim: \project increases the number of concurrently active agents by exploiting the \AG pattern, and the benefit scales with both agent count and model size.

\subsection{Why Scale Improves: Collective Reuse Compute}
\label{sec:eval-compute}

Does collective \KVC reuse amortize the reuse overhead that grows with agent count in request-centric methods?
The scaling result in Q1 combines compute and memory gains.
We now isolate the compute contribution by measuring how collective \KVC reuse affects prefill throughput as agent count grows.

We replay a single \GA round while varying the number of agents (3, 5, 10, 15, and 20) and the offered QPS (1, 2, 4, 8, 12, and 16).
All agents share the same set of output blocks from the previous round, so the degree of cross-agent redundancy is controlled.
We measure the speedup of \project's collective \KVC reuse over the serial baseline, where each request is processed by the PIC backend independently.

\begin{figure}[t]
\centering
\includegraphics[width=\columnwidth]{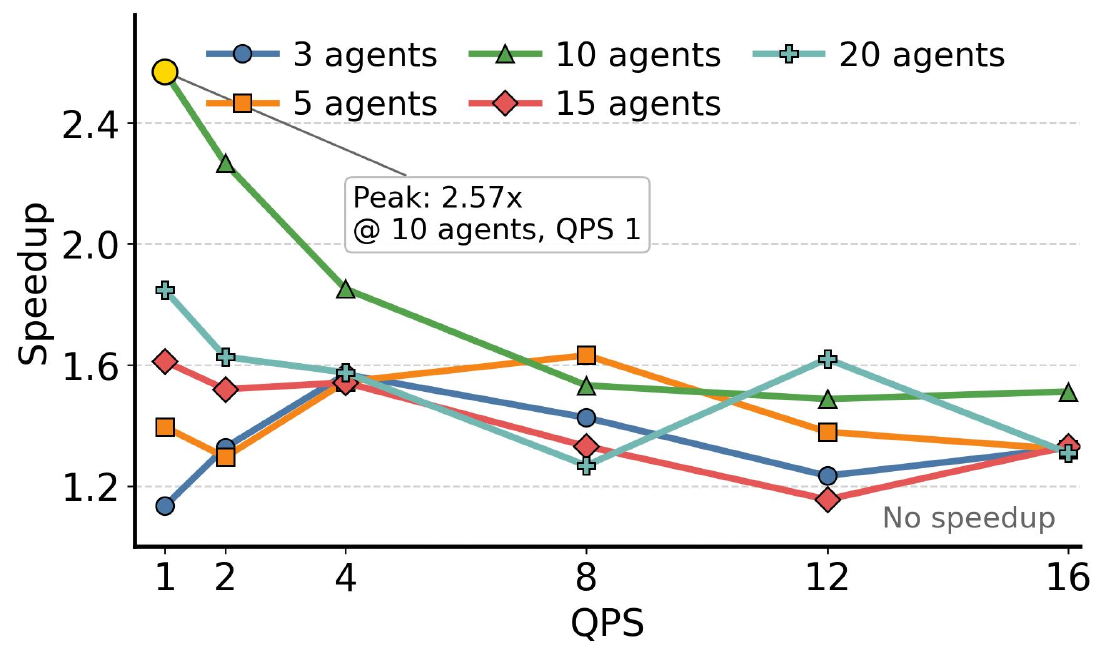}
\caption{
Collective \KVC reuse speedup over serial (per-request) PIC recovery for varying agent counts (3, 5, 10, 15, 20) and QPS levels on the \GA workload.
The peak speedup of $2.57\times$ occurs at 10 agents and $QPS = 1$.
All configurations exceed $1.0\times$ across the full QPS range, confirming that collective reuse is consistently beneficial.
}
\label{fig:exp_kv_reuse}
\end{figure}

Figure~\ref{fig:exp_kv_reuse} reports the results.
The y-axis shows the speedup of collective reuse over the serial path; a value above 1.0 means collective reuse is faster.
The peak speedup is $2.57\times$, achieved at 10 agents and $QPS = 1$.
At low QPS, larger groups benefit more from amortization: the 10-agent configuration reaches $2.57\times$ and the 5-agent configuration reaches $1.31\times$ at $QPS = 1$, while the 3-agent configuration achieves a more modest $1.17\times$.
The 15-agent and 20-agent configurations reach $1.61\times$ and $1.73\times$ at $QPS = 1$ respectively; beyond 10 agents, scheduling overhead grows and partially offsets the per-request amortization gain.

Two observations stand out.
First, all agent counts achieve speedup above 1.0 across the entire QPS range, confirming that collective reuse is consistently beneficial regardless of group size.
The benefit is largest at low QPS where the system has time to fully exploit the collective grouping, and smallest at high QPS where GPU compute saturation limits the margin.
Second, as QPS increases beyond 4, the speedup converges across agent counts to the $1.2$--$1.6\times$ range.
At high QPS, the GPU becomes compute-saturated and the bottleneck shifts from redundant reuse work to raw compute capacity.
Nonetheless, even at $QPS = 16$, collective reuse still provides a $1.3$--$1.5\times$ speedup, which directly translates to the higher agent capacity observed in Q1.

The key takeaway is that collective \KVC reuse amortizes the RoPE and important-position selection cost across the round.
The per-request overhead of reuse analysis grows sublinearly with agent count under \project, whereas it grows linearly under request-centric methods like CacheBlend.
This sublinear scaling is what shifts the latency curve downward in Figure~\ref{fig:exp_scale} and allows more agents to fit under the SLO.

\subsection{Why Scale Improves: \KVC Footprint and Storage Redundancy}
\label{sec:eval-memory}

Does Master-Mirror storage significantly lower the \KVC cost per agent round?
Collective \KVC reuse reduces compute waste, but the system still holds one recovered \KVC per agent.
If these caches remain dense and duplicated, memory pressure will still cap the number of active agents.
We now isolate the memory contribution by characterizing the redundancy among recovered caches and measuring the storage savings of the Master-Mirror layout.

\begin{figure}[t]
\centering
\includegraphics[width=\columnwidth]{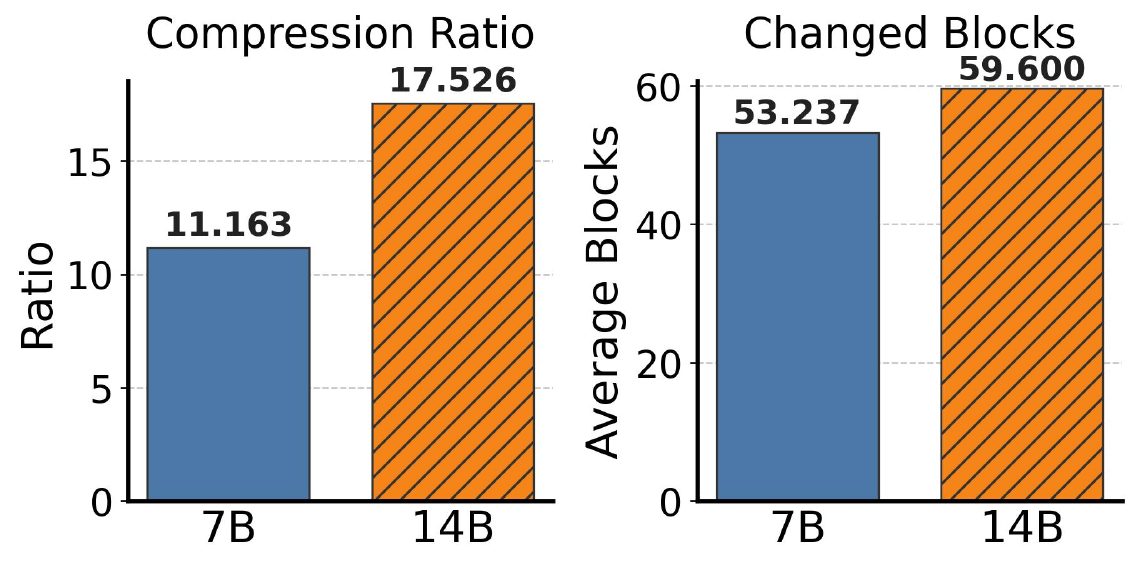}
\caption{
Redundancy characterization of recovered \KVCs across agents in a single \GA round.
Left: compression ratio (full cache size divided by Master-plus-diff size).
Right: average number of 32-token blocks that differ between a Mirror and its Master.
The 14B model achieves higher compression because cache tensors per token are larger while the number of differing blocks stays similar.
}
\label{fig:exp_redundancy}
\end{figure}

\noindent\textbf{Redundancy characterization.}
We produce recovered \KVCs for all agents in a single \GA round and compare each Mirror against the selected Master at block granularity (32 tokens per block).
Figure~\ref{fig:exp_redundancy} reports two metrics for both the 7B and 14B models.
The left panel shows the compression ratio, defined as the ratio of the full dense \KVC size to the Master-plus-diff size.
The 7B model achieves a compression ratio of $11.2\times$, meaning that the diff-encoded Mirror is roughly $9\%$ the size of a full cache.
The 14B model achieves $17.5\times$, because the larger model produces longer cache tensors per token while the number of differing blocks remains similar.
The right panel shows the average number of changed blocks per Mirror.
The 7B model averages 53.2 changed blocks per Mirror, and the 14B model averages 59.6.
These numbers are small relative to the total block count of a full cache (typically 500--700 blocks in these workloads), confirming that the vast majority of blocks are identical to the Master.

The higher compression ratio on the 14B model is significant for scaling.
Each 14B agent's full \KVC is approximately twice the size of a 7B agent's cache.
By compressing Mirrors to about $6\%$ of the full size, \project effectively removes the model-size penalty from the per-agent storage cost.
This explains why \project's scaling advantage over the baselines is larger on the 14B model in Figure~\ref{fig:exp_scale}: the storage savings translate directly into more agents fitting within the GPU memory budget.

\noindent\textbf{Implied capacity gain.}
The compression ratios above translate to concrete capacity improvements.
Under the Master-Mirror layout, the \KVC cost of $N$ agents in one round is approximately $1 + (N-1)/R$ full caches, where $R$ is the compression ratio.
For $N = 10$ agents on the 7B model ($R = 11.2$), this is $1 + 9/11.2 \approx 1.8$ full caches instead of $10$, a $5.6\times$ reduction in \KVC memory.
For the 14B model ($R = 17.5$), the cost is $1 + 9/17.5 \approx 1.5$ full caches instead of $10$, a $6.7\times$ reduction.
This memory reduction is the mechanism that allows \project to sustain more agents within the same GPU memory budget.

\subsection{End-to-End Retrieval Cost}
\label{sec:eval-fused}

Does the fused retrieval path preserve the storage gains during online serving, or does reconstruction latency erase the compression benefit?
A naive restore path would reconstruct a dense Mirror by copying the full Master and overwriting the differing blocks, adding an extra write-then-read round trip on the critical path.
We compare this dense restore approach against \project's fused diff path, which applies sparse corrections during the layerwise GPU transfer pipeline without materializing a separate dense copy.

\begin{figure}[t]
\centering
\includegraphics[width=\columnwidth]{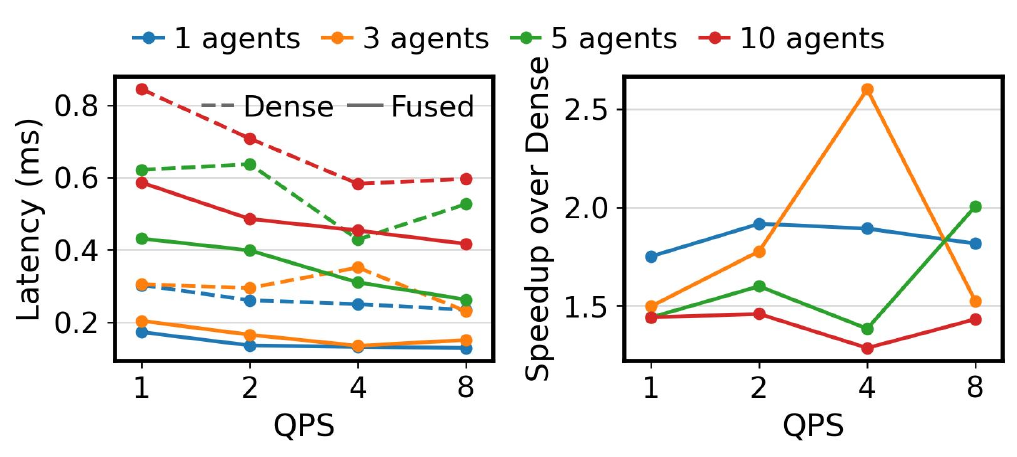}
\caption{
Latency analysis of Mirror state reconstruction on \GA using Qwen2.5-7B.
Left: absolute restore latency for dense reconstruction (dashed) and fused diff retrieval (solid) across agent counts and QPS levels.
Right: speedup of fused retrieval over dense restore.
Fused retrieval consistently reduces restore latency by $1.3$--$2.6\times$ by avoiding a separate dense materialization step.
}
\label{fig:exp_storage}
\end{figure}

Figure~\ref{fig:exp_storage} reports the results for 1, 3, 5, and 10 agents at QPS levels from 1 to 8 on the \GA/Qwen2.5-7B configuration.
The left panel plots absolute restore latency for both paths.
Dashed lines show dense restore; solid lines show fused retrieval.
At 10 agents and $QPS = 1$, dense restore takes $0.59$ ms per Mirror, while fused retrieval takes $0.43$ ms, a $27\%$ reduction.
At 1 agent, fused retrieval costs $0.13$--$0.18$ ms, which is negligible compared to the overall round latency of several hundred milliseconds.
As agent count increases, the absolute gap between dense and fused paths widens because fused retrieval avoids materializing increasingly large dense tensors.

The right panel reports the speedup of fused retrieval over dense restore.
The speedup ranges from $1.3\times$ to $2.6\times$ depending on agent count and QPS.
The 3-agent configuration achieves the highest peak speedup of $2.6\times$ at $QPS = 4$.
The 1-agent configuration shows a consistent $1.8$--$2.0\times$ speedup across the QPS range.
At 10 agents the speedup is $1.3$--$1.5\times$ because the restore pipeline becomes partially bandwidth-bound regardless of the reconstruction strategy.
Across all configurations, fused retrieval is consistently faster than dense restore, confirming that the storage compression from Q3 is preserved on the online critical path.

The combination of Q3 and Q4 closes the memory-side argument.
Master-Mirror storage reduces the per-agent \KVC footprint by $11$--$17\times$ (Q3), and fused diff retrieval ensures that accessing the compressed representation adds less latency than reconstructing a dense copy (Q4).
Together, these two mechanisms enable the higher agent counts observed in Q1.

\subsection{Accuracy Influence}
\label{sec:eval-accuracy}

Does \project's collective reuse and compressed storage alter model output compared to a standard per-request serving system under greedy decoding?
By construction, \project's collective path and Master-Mirror storage produce the same recovered \KVCs as the underlying PIC method applied per request; the collective grouping changes execution order but not the numerical result.
In our evaluation we use CacheBlend as the PIC backend, so \project produces the same output as CacheBlend with per-request recovery.
Any output difference relative to a non-PIC baseline (e.g., vLLM with prefix caching) is therefore attributable to CacheBlend's selective recomputation, not to \project.

\begin{figure}[t]
\centering
\includegraphics[width=\columnwidth]{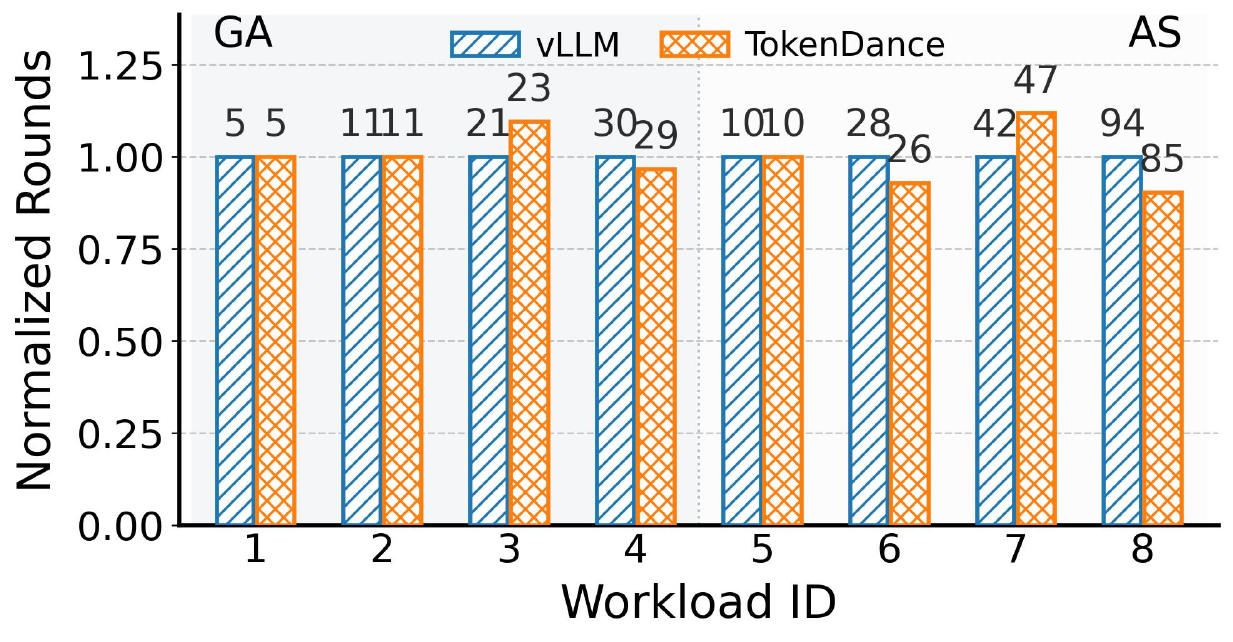}
\caption{
Accuracy evaluation across eight scenarios from \GA (IDs 1--4) and \AS (IDs 5--8).
Each bar shows the number of simulation rounds completed before the first output divergence between \project and vLLM (prefix caching, temperature~$= 0$).
The relative difference $\Delta$ is annotated above each pair.
Workload ID mapping: 1~=~Meet and Greet, 2~=~Valentine's Day Party, 3~=~Election Discussions, 4~=~Winning the Election, 5~=~Information Outbreak, 6~=~Pre-Landfall Activity, 7~=~Hurricane, 8~=~Economic Stabilization.
Three scenarios show zero divergence; the remaining differences are attributable to the underlying PIC method, not to \project.
}
\label{fig:exp_accuracy}
\end{figure}

To verify this, we run both \project and vLLM with prefix caching on the same set of agent rounds, setting temperature to $0$ to eliminate sampling randomness and using identical system prompts and input histories for every agent.
For each scenario we record the number of simulation rounds that complete before the first divergence, i.e., the first round in which at least one agent produces a different response under the two systems.
Figure~\ref{fig:exp_accuracy} reports the results across eight scenarios spanning both workloads (IDs 1--4 from \GA, IDs 5--8 from \AS).

In three of the eight scenarios (Meet and Greet, Valentine's Day Party, and Information Outbreak), both systems produce identical output for the entire trace ($\Delta = 0.0\%$).
In the remaining five scenarios, the round counts differ by $3.3\%$--$11.9\%$.
These differences arise because CacheBlend's PIC recovery introduces small numerical perturbations at selectively recomputed positions;
under greedy decoding, such perturbations can eventually flip a single token choice, which then cascades through subsequent rounds.
This is consistent with the stochastic nature of multi-agent simulation, where minor perturbations in one round propagate unpredictably through later rounds.
The key result is that \project does not introduce any additional accuracy degradation beyond what the underlying PIC method already produces.

\section{Related Work}
\label{sec:related_works}

\noindent\textbf{LLM Serving Systems.}
vLLM~\cite{vllm}, SGLang~\cite{sglang}, Orca~\cite{orca}, and Sarathi-Serve~\cite{sarathi} improve batching, scheduling, and memory management for LLM inference.
These systems optimize execution for individual requests but do not exploit the round-level \KVC redundancy that \project targets.

\noindent\textbf{Request-Centric \KVC Reuse.}
Prefix caching in SGLang~\cite{sglang} and vLLM~\cite{vllm} reuses cached state when a new request shares an exact prefix with a stored sequence.
PromptCache~\cite{promptcache} reuses cached modules identified by markup, and DroidSpeak~\cite{droidspeak} shares \KVCs across different LLMs.
EPIC~\cite{epic}, KVLink~\cite{kvlink}, and KVComm~\cite{kvcomm} recover reuse at arbitrary positions through position correction, and CacheBlend~\cite{cacheblend} further adds selective recomputation to restore accuracy at important positions.
All of these methods operate on one request at a time.
\project differs by treating the agent round as the reuse unit: it shares reuse work across all agents in a round and compresses the resulting set of \KVCs that this collective execution produces.

\noindent\textbf{\KVC Compression and Movement.}
HCache~\cite{hcache}, InfiniGen~\cite{infinigen}, CacheGen~\cite{cachegen}, MoonCake~\cite{mooncake}, and CachedAttention~\cite{cachedattention} reduce the cost of long-context serving through recomputation, offloading, compression, or disaggregation.
These methods target individual caches.
\project instead exploits similarity across sibling caches from the same round.
The Master-Mirror layout and fused diff path are designed around cross-cache redundancy rather than single-cache compression alone.

\noindent\textbf{\KVC Eviction and Quantization.}
H2O~\cite{h2o} and ScissorHands~\cite{Scissorhands} reduce \KVC memory by evicting tokens with low attention scores, while StreamingLLM~\cite{streamllm} and FastGen~\cite{modeltellsyouwhattodiscard} keep only a fixed window plus a few important tokens for long-sequence serving.
KIVI~\cite{kivi} and KVQuant~\cite{kvquant} take a different path, quantizing cached values to 2--3 bits with little accuracy loss.
These methods shrink individual caches but do not address cross-cache redundancy.
\project is orthogonal: it removes the duplicated data across sibling caches first, and per-cache compression can be applied on top.

\noindent\textbf{Attention Kernels and Sparse Backends.}
FlashAttention~\cite{flashattention} shows that attention performance depends heavily on data movement.
FlashInfer~\cite{flashinfer} provides efficient variable-length and block-sparse attention backends.
Our fused diff kernel follows the same data-movement logic.
It aligns sparse diff blocks with attention tiles so compressed caches can be used without an extra dense reconstruction pass.

\noindent\textbf{Multi-Agent LLM Systems.}
Parrot~\cite{parrot}, Autellix~\cite{autellix}, Teola~\cite{teola}, ScaleSim~\cite{scalesim}, and Tokencake~\cite{tokencake} optimize scheduling, prefetching, or execution order for agent applications.
LRAgent~\cite{lragent} shares \KVCs across multi-LoRA agents by decomposing caches into shared and adapter-specific parts.
\GA~\cite{generativeagents}, \AS~\cite{agentsociety}, OpenClaw~\cite{openclaw}, and MoltBook~\cite{moltbook} show that many real agent workloads run in explicit rounds with a shared scheduler.
These systems define or schedule agent applications; \project optimizes the \KVC layer for the repeated gather-and-redistribute pattern that they all share.

\section{Conclusion}
\label{sec:conclusion}

We presented \project, a system for scaling multi-agent LLM serving through collective \KVC sharing.
The design follows two core observations.
Current reuse methods are inefficient for agent rounds, and even after round-level \KVC reuse the resulting \KVCs remain highly compressible.
\project turns these observations into a progressive design with collective \KVC reuse, Master-Mirror storage, and fused diff retrieval.
On representative multi-agent workloads, it reduces end-to-end latency by up to $2.3\times$ and \KVC storage by $94\%$ compared to vLLM with prefix caching.

More broadly, our work argues that communication structure should become a first-class concept in LLM serving.
The \AG pattern is a natural starting point because it already appears across many agent workloads.
As agent systems continue to diversify, other recurring communication patterns will likely expose similar opportunities.
A communication-pattern-aware serving stack will be important for making future multi-agent platforms efficient at scale.

\bibliography{reference}

\end{document}